\documentclass[nofootinbib,floatfix,aps,pra]{revtex4}
\usepackage{graphicx}
\usepackage{color}
\usepackage{verbatim}
\usepackage{threeparttable}
\usepackage{amsmath,amsfonts}
\newcommand{\be}{\begin{equation}}
\newcommand{\ee}{\end{equation}}
\newcommand{\bea}{\begin{eqnarray}}
\newcommand{\eea}{\end{eqnarray}}

\begin{document}
\title{Accurate Anisotropic Gaussian Type Orbital Basis Sets for 
Atoms in Strong Magnetic Fields} 

\author{Wuming Zhu,$^{1,}$\footnote{\label{email} Electronic addresses: 
zhuwm@hznu.edu.cn and trickey@qtp.ufl.edu} and S.B.\ Trickey$^{2,\ref{email}}$}

\affiliation{$^1$Department of Physics, 
Hangzhou Normal University, 
16 Xuelin Street, Hangzhou, Zhejiang 310036, China}

\affiliation{$^2$Quantum Theory Project,
Dept.\ of Physics and Dept.\ of Chemistry,
P.O.\ Box 118435, University of Florida, Gainesville FL 32611-8435 U.S.A.}

\begin{abstract}
In high magnetic field calculations, 
anisotropic Gaussian type orbital (AGTO) basis functions
are capable of reconciling the competing demands of the spherically 
symmetric Coulombic interaction and cylindrical
magnetic ($B$ field) confinement. However, the best available {\it a priori} 
procedure for composing highly accurate  AGTO sets for atoms 
in a strong $B$ field [Phys.\ Rev.
 A {\bf 90}, 022504 (2014)] yields very large basis sets.  Their 
size is problematical for use in any calculation with unfavorable 
computational cost scaling.    Here we provide an
alternative constructive procedure.  It is based upon   
analysis of the underlying physics of atoms in $B$ fields
that allows identification of several principles for the construction 
of AGTO basis sets. 
Aided by numerical optimization and parameter fitting, 
followed by fine tuning of fitting parameters, 
we devise formulae for generating accurate AGTO basis sets 
in an arbitrary $B$ field. For the hydrogen iso-electronic sequence, 
a set depends on $B$ field strength,  nuclear charge, and upon 
orbital quantum numbers. 
 For multi-electron systems, the basis set formulae also include 
adjustment to account for orbital occupations. 
Tests of the new basis sets for atoms H through C ($1 \le Z \le 6$), 
and ions Li$^+$,  Be$^+$,  and B$^+$, in a wide $B$ field range ($0 \le B \le 2000$ a.u.), 
show an accuracy better than a few $\mu$H for single-electron systems, and a few
hundredths to a few mHs for multi-electron atoms.
The relative errors are similar for different atoms and ions in a large  
$B$ field range, from a few to a couple of tens of millionths, thereby 
confirming rather uniform accuracy across the nuclear charge $Z$ 
and $B$ field strength values.  
Residual basis set errors are two to three orders of magnitude smaller than the
electronic correlation energies in muti-electron atoms, a signal of 
the usefulness of the new AGTO basis sets in correlated wavefunction 
or density functional calculations
for atomic and molecular systems in an external strong $B$ field. 
\end{abstract}
\pacs{}
\maketitle


\section{\label{Intro}Introduction}

Theoretical calculations for atoms and molecules subjected to a strong
magnetic field ($\mathbf B$) are indispensable to the interpretation
of observed spectra and to ascertainment of the elemental composition
of the atmosphere of those celestial bodies which exist in a strong
B-field, e.g., the fields of magnetic white dwarfs ($10^2-10^5$ T) and
neutron stars ($> 10^7$ T).  Although such large field strengths far
exceed terrestrially available values, typical strong field effects
nevertheless can be observed in laboratories in which static fields
up to $30-40$ T can be generated. Example systems include Rydberg
atoms and the exciton in a semiconductor which possesses a small
effective mass and a large dielectric constant \cite{Chui-PRB-1974}.

A non-perturbative approach is required to address the problem of atoms and 
molecules in a strong B field because the Coulomb force and the Lorentz force
experienced by the electrons are on a par. A variety of methods thus has been 
devised to cope with the problem. They include formal mathematical 
analysis \cite{LiebBook}, algebraic methods \cite{Liberman96PRA}, 
representation of the electronic wave function on a two-dimensional
mesh \cite{Ivanov, Heyl}, or as a linear expansion in a suitably chosen basis.
Basis sets used include finite-element schemes \cite{Braun}, B-splines \cite{ZhaSta-PRA-2008}, Slater-type orbitals (STO) \cite{Ceperley96PRA} 
(which are more suitable to the weak B regime), Landau functions 
\cite{Relovsky, MedLai-PRA-2006} (which are suitable for the very 
strong B field regime in which the adiabatic approximation 
is applicable \cite{SchiffSnyder39}), or the product of both 
\cite{CheGol-PRA-1992}.  
 
The medium to high field strength regime poses a formidable challenge
to the calculations because of different symmetries of Coulomb and
Lorentz interactions, neither of which is treatable as a
perturbation. Current best practice for this difficult regime is to
employ an anisotropic Gaussian type orbital (AGTO) basis.  In such a 
basis, the exponential factor of the basis functions has different
decay constants along the direction of the B field and the direction
perpendicular to it.  This anisotropy provides the flexibility
required to describe elongation of electron orbitals and densities
along the field direction.  

AGTO basis sets were first introduced, to our knowledge, by Aldrich
and Greene \cite{Aldrich}, and exploited extensively by Schmelcher and
Cederbaum \cite{Schmelcher88PRA}.  Successful applications are found
in several studies of atoms in strong $B$ fields
\cite{Becken,Ceperley99PRA,Al-Hujaj04Li,Al-Hujaj04Be,WanQia-PRA-2007}.
As with ordinary GTO basis sets, the lack of a unique formal
prescription for determination of the basis exponents compels
development of a physically sound prescription.  Kappes and Schmelcher
gave an optimization algorithm \cite{Kappes}.  Because it must be used
for every distinct combination of atomic configuration and field
strength, computational cost and tedium are substantially larger than
with an ordinary ground-state GTO set.  Furthermore, a good starting
set of exponents is required \cite{Becken}.  Consequently, it is
almost dauntingly hard and computationally costly to achieve near
uniformity of quality of the generated basis sets.  The magnitude of
the residual basis set errors is uncontrolled, with no guarantee that
those errors are insignificant nor close enough to uniform to support
unbiased comparisons between different atomic configurations or
different field strengths.

An alternative to case-by-case optimization is the
use of nearly optimized basis sets. Kravchenko and Liberman (KL)
\cite{Liberman97IJQC} investigated one-electron systems, the hydrogen
atom and the hydrogen molecular ion. They found that systematically
constructed AGTO basis sets could provide absolute accuracy (relative
to exact solution) of $10^{-6}$
Hartree or better. 
We later improved the construction procedure and produced a
set of formulae to generate nearly optimized basis sets based on a simple
physical model. Details can be found in
Ref.\ \cite{ZhuZhaTri-PRA-2014}.  Unfortunately, both the KL
basis sets and ours rely upon multiple sequences of exponents (up through
five) to reach relatively high accuracy, especially for large $B$.  
The inescapable result is very large basis set size when compared to 
ordinary high-quality ground-state GTO sets.  Such large sizes are 
quite unfavorable for use in highly correlated method 
calculations. Those methods typically have rather severe  
computational cost scaling with basis set cardinality (some 
power higher than four of basis set size times number of electrons). 
Therefore it is highly desirable to keep the basis set size as small
as possible commensurate with desired accuracy.  Experience with
our multiple-sequence sets strongly suggests that we should have only 
one sequence of basis exponents, or at most two if that is unavoidable.

For the $B=0$ case,  a huge variety of well-developed basis sets 
is available from which to choose. Among them, two with very long
lineages in quantum chemistry (beginning in the 1960s) are the 
Pople-type sets which use minimal representations of  Slater-type 
functions or split-valence
primitives to describe core and valence electron wavefunctions
\cite{PopleBasis}, and the Dunning correlation-consistent
basis sets. The latter sets are composed of Gaussian
primitives contrived to provide 
systematic convergence of the correlation
energy from post-Hartree-Fock (HF) calculations \cite{DunningBasis}.  
For $B>0$, there is no obvious 
single path to generalizing either the Pople or Dunning type sets because
of the infinite number of combinations of atomic nuclear charge $Z$
and $B$ field strength.  Instead, one must seek appropriate
formulations as a function of field strength and, as we discuss
below, orbital type as helpful parameters. In doing so, 
we follow the
philosophy adopted by Dunning \cite{DunningBasis}. We explore
(analytically and numerically) some simple 
paradigmatic systems to establish a few principles which 
should be applicable in general. We try to fathom, or at least
rationalize, the underlying physical reasons
that dictate those principles.  We do some fine-tuning and test
the resulting formulas for basis set generation with atoms H through C ($1 \le Z
\le 6$, where $Z$ = atomic number) over a very large range of $B$
field strength ($0 \le B \le 2000$ a.u.).  We take advantage of the
availability of several highly accurate HF calculations
\cite{Ivanov,Heyl,ZhaSta-PRA-2008,Ceperley96PRA}
to assess the residual errors associated with the new basis set construction
formulation. 

Density functional theory (DFT) calculations are not included in this
study because the extra errors originating from numerical grids and
variety of exchange-correlation functionals would complicate 
comparison between our calculated results and reference values.  
Nevertheless, the basis set construction procedure presented
here is applicable to
DFT calculations as well as to wavefunction based methods.

The next section gives the essential details of the problem and
reviews our previous basis set construction
procedure. \cite{ZhuZhaTri-PRA-2014}. We single out the points that
should be kept in the new procedure and discern principles and
requirements omitted.  Section \ref{HBasis} details the basis set
construction for the
hydrogen iso-electronic sequence in an arbitrary $B$ field. 
From the newly constructed one-electron system basis set, 
in subsection \ref{ATMBasis} we extend the treatment to multi-electron
atoms or atomic ions by modification 
to recognize orbital occupations.  
The result is a set of formulae for constructing basis sets for 
multi-electron atomic systems
in a $B$ field. In section \ref{Concl}, we summarize and make a 
few conclusions.

\section{\label{Theo}Methodology}
\subsection{\label{Basics}Basics}

The system Hamiltonian  of a central field atom at the origin 
in a uniform, static 
external magnetic field along the z direction,
${\mathbf B} = B \hat z $, 
commutes with rotations about $\hat z$, 
so the magnetic quantum number $m_i$ of the {\it i}th electron 
remains good.
The total spin ${\bf S}^2$,
its $z$ component $S_z$, and spatial parity in $z$ also are
conserved, so a state 
can be labeled by those quantum numbers. We choose the 
Coulomb gauge,  
${\mathbf A}({\mathbf r})=\frac {1}{2} {\mathbf B} \times {\mathbf r}$,
with ${\mathbf A}$ the vector potential.  
Then the system Hamiltonian is
\be
{\bf {\it H}}=\sum_{i}\left [ - \frac{1}{2} {\nabla}_i^2 - \frac{Z}{r_i} %
+ \frac{B^2}{8} \left (x_i^2 + y_i^2 \right)  + \frac{B}{2}  \left (m_i + 2m_{s,i} \right) \right ] %
+\frac {1}{2} \sum_{i \neq j}\frac {1}{\left |{\bf r}_i-{\bf r}_j \right |}  \; ,
\label{Hamiltonian}
\ee
with ${\mathbf r}_i$ and $m_{s,i}$  the space coordinate, 
and spin $z$ component for the ${\it i}$-th electron.  (Hartree atomic units 
($\hbar = m_{electron} = q_{electron} =1 $) are used throughout  
unless otherwise noted. One Hartree a.u.\ of magnetic field
equals $2.3505 \times 10^5$ Tesla.)
In what follows, unpaired electrons are always taken 
as spin down, $m_{s,i} = - \frac{1}{2}$ unless stated otherwise.   

\subsection{\label{Previous}Previous Construction Scheme}
In cylindrical coordinates $(\rho, z, \phi)$,  the ${\it j}$-th 
AGTO basis function takes the form
\be
\chi_j(\rho,z,\phi) = N_j \rho^{n_{\rho_j}} z^{n_{z_j}} e^{-\alpha_j \rho^2 - \beta_j z^2} %
e^{i m_j \phi},    j = 1,2,3,\ldots
\label{basis}
\ee
where $n_{\rho_j} = |m_j| + 2k_j, k_j=0,1,\ldots$, with $m_j = \ldots,-2,-1,0,1,2,\ldots$,
and   $n_{z_j}  = \pi_{z_j} + 2l_j, l_j=0,1,\ldots$, with $\pi_{z_j} = 0,1.$
Such functions are advantageous for the study of atoms in strong
$B$ fields because of their flexibility in describing the elongation of the
electron orbitals and densities along the $B$ field direction.  
Since the $B$ field compresses the transverse (radial) functions 
(related to exponents $\alpha_j$) 
relative to the corresponding longitudinal (axial) ones (exponents $\beta_j$),
we always have $\alpha_j \ge \beta_j$.
Schmelcher and Cederbaum deduced all the required matrix elements of 
the Hamiltonian with respect to such AGTO basis function 
\cite{Schmelcher88PRA}. 
However, the optimal determination of the sets  $\lbrace \alpha_j \rbrace$
and $\lbrace \beta_j \rbrace$ was left largely as an open question.   

As already mentioned previously, we constructed 
KL-like highly optimized basis sets 
without requiring case-by-case full non-linear optimization \cite{ZhuZhaTri-PRA-2014}.  We summarize that procedure to lay the ground work for the
new scheme.  

A very highly accurate total energy for the non-relativistic
H atom at $B=0$ a.u.\ is given by an even-tempered Gaussian (ETG)
sequence of length $N_b$ and rule of formation \cite{SchmidtRuedenberg},   
\bea
\beta_j & = & pq^j \; , \; j=1,2,\ldots,N_b \nonumber \\
\ln p & = & a \ln (q-1) + a^\prime \nonumber \\
\ln(\ln q) &=& b \ln N_b + b^\prime \nonumber \\
a &=&0.3243\, , \;\;\; a^\prime = -3.6920 \nonumber \\
b &=&-0.4250\, , \;\;\; b^\prime = 0.9280 
\label{eventemp}
\eea
with $N_b = 16$. This 
%
determines the parameters $p$ and $q$, hence the longitudinal exponents $\beta_j$, of what we choose as the ``base sequence'', that is, the exponent
set with reference to which all the others are constructed.  

First, since the magnetic field does not change the confinement along 
$z$, we expect little effect of $B$ upon the ETG exponents.  Therefore 
our construction 
retained the ETG sequence Eq.\ (\ref{eventemp}) unchanged for the longitudinal 
exponents $\beta_j$.  The one modification occurs in cases for which 
the electron density has diffuse, non-zero orbital angular momentum 
contributions.  Those require that the tempering be extrapolated 
to include a small number of negative
$j$ values.  

Then, by consideration of large and small $B$ field limits and use
of a nonlinear fit to calculated one-electron system results, Ref.\ 
\cite{ZhuZhaTri-PRA-2014} 
reached a prescription for highly refined, nearly optimal exponents 
for Eq.\ (\ref{basis}). 
Those  AGTO basis sets are comprised of one to five exponent sequences 
($\ell = 1,2,3,4,5$), 
Each sequence is a subset of exponents related by 
\bea 
\alpha_{j,\ell} &  = & \beta_j + (1+ \mu_\ell) \Delta(B) \label{alphas} \\
\mu_1 & = & 0.0 \; , \; \mu_{2, 3} = \pm 0.2 \; , \; %
\; \mu_{4, 5} = \pm 0.4 \; ,
\label{BasSeqs}
\eea
and
\bea
\Delta(B) = 
\frac {B}{20} \left \{ 4 \left [ 1 + \frac {4}{b(\gamma)} \frac {\beta_j}{B} \right ]^{-2} %
+ \left [ 1 + \frac {4}{b(\gamma)} \frac {\beta_j}{B} \right ]^{-1/2} \right \}
\label{DeltaparenB}
\eea
where 
\bea
b(\gamma) &=& -0.16 [tan^{-1}(\gamma)]^2 + 0.77 tan^{-1}(\gamma) + 0.74 \; ,
\label{b-gamma} \\
\gamma &=& B/Z^2  \;.
\label{gamma-defn}  
\eea
The initial sequence for these exponents 
is $\mu_{\ell = 1} = 0$.  Then 
$\mu_\ell = \pm 0.2, \;  \pm 0.4$ for the second,
third, fourth, and fifth sequences, respectively.  For 
$\ell =2$, $3$, there are half as many exponents (with doubled
spacing) as in the base sequence, while for 
$\ell = 4$, $5$ there are one-fourth as many (with quadrupled spacing). 

Testing showed that the resulting basis sets, Eqs.\ (\ref{alphas}) - (\ref{gamma-defn}), 
introduce errors of less than $1 \mu$Hartree for the H atom 
over $0 \le B \le 2000$ a.u.\ as compared with  
more accurate algebraic results \cite{Liberman96PRA}. 
Further, the HF total energies for multi-electron atoms
from the AGTO prescription, Eqs.\ (\ref{alphas}) - (\ref{b-gamma}),  
are nearly indistinguishable from full numerical 
two-dimensional mesh results \cite{Ivanov}. 

However, our previous scheme has several limitations. 
First, the scheme makes no distinction among orbitals of different
types (different quantum numbers), yet clearly the physics of magnetic
confinement is not identical.  
Second, no consideration of the strong electron-electron interaction 
in a  multi-electron system was involved.  
Lastly, up to five sequences from Eq.\ (\ref{BasSeqs})  
were required to reach appropriate accuracy. 
Including multiple sequences may offset, in part at least, the aforementioned 
two deficiencies in the constructed basis sets.  But as already
noted, multiple sequences introduce unfavorable computational cost scaling,
especially for highly correlated methods applied to many-electron systems.
Thus the endeavor here is to include as few sequences as possible in our basis sets,
ideally only one.

One other aspect of our previous basis set construction scheme
bears mentioning because it is important enough that we retained
it.  This is a pair of constraints 
on $\Delta(B)$ for very small and very large longitudinal basis
set exponents.  Specifically, we take $\Delta(B)
= \frac{B}{4}$ in the limit $\beta_j \rightarrow 0$. This can be
viewed as the large $B$ limit or the zero nuclear charge limit.  
Either way, the one-electron state is a Landau orbital, with an
exponential parameter ${\alpha_j} \equiv a_B= B/4$ ($a_B$ is the scale
parameter in the Landau orbital).  The opposite limit, $\beta_j
\rightarrow \infty$, corresponds to $B = 0$ a.u. which restores spherical
symmetry, $\alpha_j = \beta_j$, hence $\Delta(B=0) = 0$. 
We found that a convenient choice for orbital exponent
asphericity is a scaling of $a_B$.  Then $\Delta(B)$ is taken as a
monotonically decreasing function of the increasing longitudinal
exponents $\beta_j$ of the functions in a given basis set. The
underlying physics is that a small-exponent basis function samples a
large volume far from the nucleus. In that region the $B$ field 
far outweighs the  
nuclear attraction, with the consequence that the distortion 
(relative to the field-free
spherical symmetry) will be relatively larger than for the region
sampled by larger-exponent basis functions.

\section{\label{New}New Construction Scheme}

Extension of Eqs.\ (\ref{alphas}) to (\ref{b-gamma}) 
is required to allow better
fitting to the numerically optimized transverse exponents.
But this must be done with a concurrent goal of simplification,
namely getting rid of the multiple sequences.  Substantial 
numerical experimentation and testing of several plausible 
expressions was required to arrive at expressions that 
seem superior because of their combined 
simplicity, accuracy, and flexibility.  The easiest way to 
understand the rationale is to recognize that the construction 
has two phases, one-electron and multi-electron.  The next
two subsections present those. 

\subsection{\label{HBasis}Basis Sets for Hydrogen atom and Hydrogen-like Ions in $B$ field}

Since there is only one electron in 
the hydrogen iso-electronic sequence, electron-electron interaction
disappears in these paradigmatic systems.  The first step is to 
optimize transverse exponents $\alpha_j$.  Unlike our previous study, in
which we optimized the $\alpha_j$ by starting from the 
spherical base sequence ($\alpha_j = \beta_j$), Eq.\ (\ref{eventemp}) 
and used a down-hill algorithm \cite{ZhuZhaTri-PRA-2014}, in
this study we started from a base sequence generated by
Eq.\ (\ref{DeltaparenB}). Simply because of its explicit asphericity,
it undoubtedly is an improved starting point.  Then we employed 
conjugate gradient algorithm searching in the
parameter space $\{ \alpha_j \}$ to minimize the total energy of the
system.  That produces both faster and more reliable numerical convergence
\cite{PressNRbook}.

\begin{figure} 
\resizebox{0.9\linewidth}{0.7\linewidth}{\includegraphics{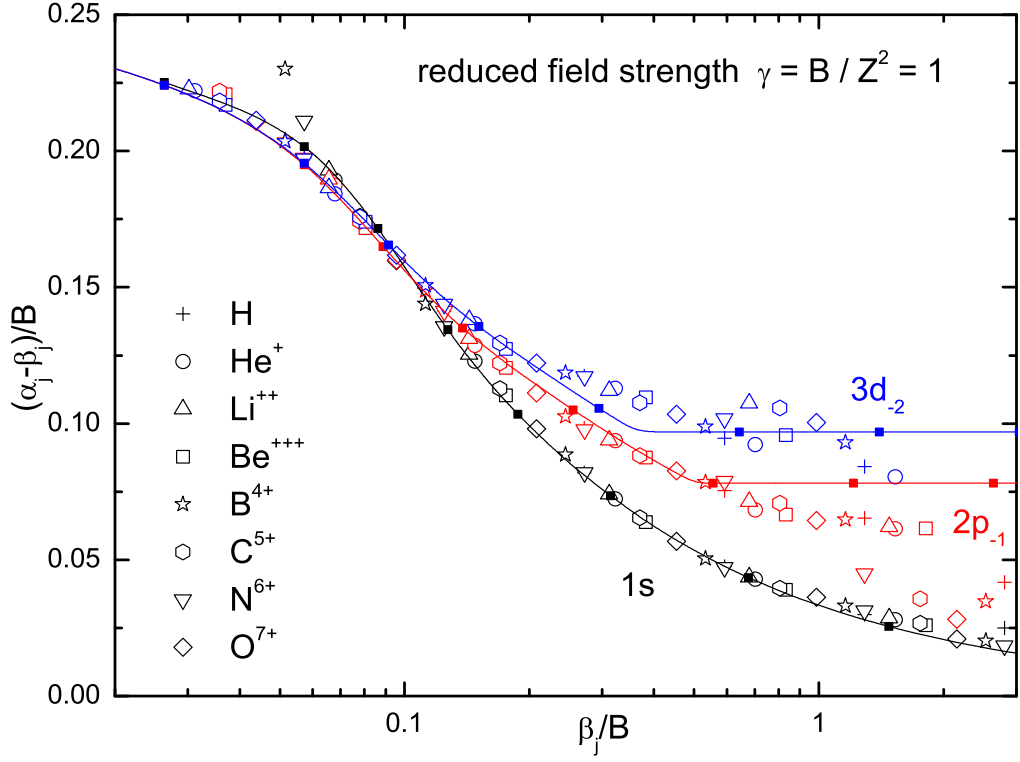}}
\vspace*{5cm}
\caption{(Color online) Fitting the optimized exponents of AGTO basis functions 
for hydrogen atom and hydrogen-like ions in reduced magnetic field strength $\gamma = 1$.
Empty symbols are optimized exponents from searching  $\{ \alpha_j \}$ space. 
Solid curves are fits to those data points, generated by 
Eqs.\ (\ref{DeltaNew}) through (\ref{AlphajNew}).
Filled squares are the AGTO basis sets for the hydrogen atom in $B = 1 a.u.$ 
generated according to Eqs.\ (\ref{DeltaNew}) through (\ref{BetajNew}).
Black, red, blue colors stand for the $1s$, $2p_{-1}$, $3d_{-2}$ orbitals, respectively.}
\label{ZhuFig1}
\end{figure}

The transverse exponent ${\alpha_j}$ optimization was done for the H
iso-electronic series (H, He$^+$, Li$^{++}$, Be$^{+++}$, B$^{4+}$, C$^{5+}$, 
N$^{6+}$, and O$^{7+}$) in reduced fields $\gamma$ = 0.1, 0.2, 0.5, 1, 2, 5, 10, 20, 
50, 100, 200, 500, 1000.
Our previous scheme fit only to optimized ${\alpha_j}$'s for the $1s$ orbital 
($m = \pi_{z} = 0$), and applied the resulting expression, 
Eq.\ (\ref{DeltaparenB}), to other orbitals as well. Closer
examination, however, shows differences among
 orbitals which must be taken into account if a truly highly accurate
basis is to be had. An example of the orbital dependence is  shown in Fig.\ \ref{ZhuFig1}.  
It displays  the results of optimization for $\gamma$ = 1 
for three different orbitals, namely $1s$, $2p_{-1}$, and $3d_{-2}$.
Note the asphericity of the basis function as measured by $(\alpha_j - \beta_j) / B$,
which is the vertical axis in that plot. Several observations ensue 
from examining Fig.\ \ref{ZhuFig1}.

Firstly, the data points of a given orbital with different nuclear charges
but for fixed $\gamma$ 
fall on the same curve. For example, all the black symbols are for $1s$ orbitals,
but the nuclear charge $Z$ ranges from 1 (neutral H), to 8  (O$^{7+}$),
yet obviously they all lie on a single curve. A simple argument shows 
that this must be the case. 
Suppose the H atom wavefunction in an external $B$ field is 
$\Psi_{H}({\bf r},B)$, then Eq.\ (\ref{Hamiltonian})gives 
\begin{equation*}
\left [ - \frac{1}{2} {\nabla}^2 - \frac{1}{r} + 
\frac{B^2}{8} \left (x^2 + y^2 \right)  + \frac{B}{2}  \left (m + 2m_{s} \right) \right ] %
\Psi_{H}({\bf r},B) = E_H \Psi_{H}({\bf r},B) 
\end{equation*}
Scaling ${\bf r} \rightarrow Z{\bf r}$ 
and multiplication by $Z^2$ gives 
\begin{equation*}
\left [ - \frac{1}{2} {\nabla}^2 - \frac{Z}{r} + 
\frac{(Z^2B)^2}{8} \left (x^2 + y^2 \right)  + \frac{Z^2B}{2}  \left (m + 2m_{s} \right) \right ] %
\Psi_{H}(Z{\bf r},B) = (Z^2E_H) \Psi_{H}(Z{\bf r},B) 
\label{scaledSEq}
\end{equation*}
The scaled  Hamiltonian is the same as that of a hydrogen-like atom with nuclear
charge Z in an external field $B^\prime = Z^2B$, or equivalently, $\gamma = \frac{B^\prime}{Z^2} = B$. 
The scaled hydrogen-atom wavefunction is precisely the eigenfunction of this Hamiltonian 
with energy of $Z^2E_H$. Now we expand $\Psi_{H}({\bf r},B)$ in the optimized basis set,
\begin{equation}
\Psi_{H}({\bf r},B) = \sum_{j} a_j \chi_j 
= \sum_{j} a_j N_j \rho^{n_{\rho_j}} z^{n_{z_j}} e^{-\alpha_j \rho^2 - \beta_j z^2} e^{i m \phi} \; .
\label{PsiH}
\end{equation}
The scaled wavefunction accordingly is
\begin{equation*}
\Psi_{Z}({\bf r},Z^2 B) = \Psi_{H}(Z{\bf r},B)
= \sum_{j} a_j N_j Z^{n_{\rho_j}+ n_{z_j}} \rho^{n_{\rho_j}} z^{n_{z_j}} e^{-\alpha^\prime_j \rho^2 - \beta^\prime_j z^2} e^{i m \phi}
\end{equation*}
where $\alpha^\prime_j = Z^2 \alpha_j$, $\beta^\prime_j = Z^2 \beta_j$.
Obviously, $\frac{\alpha^\prime_j - \beta^\prime_j}{B^\prime} = \frac{\alpha_j - \beta_j}{B}$
and $\frac{\beta^\prime_j}{B^\prime} = \frac{\beta_j}{B}$, which shows that 
the two cases indeed are on the same curve.

Secondly, though the optimized data points for orbitals with different quantum numbers 
do not fall on the same curve, their trends as functions of $B$ are similar. 
To accommodate this behavior, we introduce two parameters 
that depend upon the reduced field strength and 
orbital quantum numbers, $A(\gamma,m,\pi_z)$ and $D(\gamma,m,\pi_z)$.
Their purpose is to provide orbital-dependent asphericity in 
the basis functions. 
After  extensive numerical exploration, we reached a prescription 
for the orbital-dependent version of $\Delta$, namely 
\bea
\Delta_j(\beta_j, B) = 
B \left \{ \left ( \frac{1}{4} - \frac {\beta_j}{B} \right ) \left [ 1 - \left ( 1 - e ^{-30\beta_j/B} \right )^8 \right ]
+ \frac{A(\gamma,m,\pi_z)}{(\beta_j/B)^{D(\gamma,m,\pi_z)}} \left ( 1 - e ^{-30\beta_j/B} \right )^8 \right \}
\label{DeltaNew}
\eea
Initially, the parameters $A(\gamma,m,\pi_z)$ and $D(\gamma,m,\pi_z)$ 
were obtained 
from fits to the data points such as those displayed in Fig.\ \ref{ZhuFig1} 
by use of Eq.\ (\ref{DeltaNew}) for each 
$\gamma$ = 0.1, 0.2, 0.5, 1, 2, 5, 10, 20, 50, 100, 200, 500, 1000,
and for each of five orbitals, $1s$, $2p_{0}$, $2p_{-1}$, $3d_{-1}$, and $3d_{-2}$. 
With that set of parameters $A(\gamma,m,\pi_z)$ and $D(\gamma,m,\pi_z)$ 
determined, a second fit was made  
to express those parameters as analytical functions
of reduced field strength $\gamma$ and orbital quantum numbers $m$ and $\pi_z$.
The final result is,
\bea 
D(\gamma,m,\pi_z) & = & 0.4 + \frac{0.6(l+1)/(l^2+l+1)}{1+1.105(l+1)^3\gamma^{0.425(l+2)}} \nonumber \\
A(\gamma,m,\pi_z) & = & \frac{0.02073 + 0.00035( 2\pi_z + l(l-1)/3)}{D(\gamma,m,\pi_z)^{1.25}}
\label{DeltaNewPara}
\eea
where $l = |m| + \pi_z$.

The final observation is specific to 
the H iso-electronic series.  As one sees in Fig.\ \ref{ZhuFig1}, 
except for the $1s$ orbital, the asphericity of basis functions 
does not vanish for very
tight functions when $\beta_j \rightarrow \infty$.  Instead the
higher orbital asphericities go to some non-zero, nearly constant values.
This behavior is understandable 
from decomposition of the wavefunction into radial and angular parts. 
The kinetic energy operator in the Hamiltonian Eq.\ (\ref{Hamiltonian}) acting 
on the angular part of the wavefunction  gives rise to a repulsive 
centrifugal potential for the non-zero angular momentum orbitals.  
That kills the singularity in the nuclear attractive potential.
Thus, the overpowering Coulomb attraction exists only for the $s$ orbital 
for very tight functions, for which the asphericity of $s$ basis functions, $(\alpha_j - \beta_j) / B$,
does go to zero when $\beta_j \rightarrow \infty$. 
For non-zero angular momentum orbitals, the strong repulsive 
centrifugal potential causes the coefficients $a_j$ in
the wavefunction expansion, Eq.\ (\ref{PsiH}), to become  
negligible quickly with increasing $\beta_j$ (tighter and tighter 
basis functions).  Thus it is useful and effective to use a cutoff 
as a simple approximation
to the limit of $\Delta_j(B)$ as $\beta_j \rightarrow \infty$.
We impose a minimum value for $\Delta_j(B)$ according to 
\begin{equation}
\Delta_{min}(B) =
\begin{cases}
0, & \text{if } m = \pi_z = 0,\\
\frac{0.1562 B}{1 + \gamma^{-0.55}}, & \text{if } |m| = 1 \text{and } \pi_z=0,\\
\frac{0.1744 B}{1 + 0.8 \gamma^{-0.55}}, & \text{otherwise}.
\end{cases}
\label{DeltaMin}
\end{equation}
Putting Eqs.\ (\ref{DeltaNew}) and (\ref{DeltaMin}) together
yields the highly 
optimized transverse exponents $\alpha_j$ of the AGTO basis functions for  
the hydrogen iso-electronic sequence in an arbitrary $B$ field,
\be
\alpha_{j}  =  \beta_j + \max ( \Delta_j(\beta_j, B), \Delta_{min}(B) )
\label{AlphajNew}
\ee

To this point, we have only explored the parameter space $\{ \alpha_j \}$
with $\{ \beta_j \}$
kept unchanged as in the base ETG sequence, see Eq.\ (\ref{eventemp}).
This strategy is not entirely advantageous because the basis functions which have intermediate
asphericity ($0.05 B \lesssim \alpha_j - \beta_j \lesssim 0.20 B$ )
may not be adequate in number.  
Thus we require the asphericities of any two adjacent basis functions,
say the $j$th and  $(j+1)$th, to differ by no more than $0.03 B$. 
This is a heuristic choice which seems to be well-balanced.  Its practical
effect is evident.  For any AGTO basis function $0 \le \alpha_j - \beta_j \le 0.25 B$, 
setting the maximum interval at $0.03 B$ results in there being at 
least $\frac{0.25}{0.03} \approx 8$ basis functions included for the range
over which  the asphericity changes from zero to its maximum value, 0.25. 
That is, we use a denser set of basis functions in the exponent range 
through which the basis function asphericity  changes most rapidly.
To avoid approximate linear dependencies from 
excessively dense spacing of functions, 
we also require that the ratio of the longitudinal exponents of any 
two adjacent 
basis functions be no less than $\sqrt q$, where $q$ is from Eq.\ (\ref{eventemp}). 
This is a modification of that equation. At the ratio $\sqrt q$, there 
is twice the basis function density than from use of the ratio $q$.
Heuristically this works well in avoiding approximate linear dependencies
while handling changing asphericity. 
To enforce those constraints, first we need the inverse of Eq.\ (\ref{DeltaNew}),
\be
\beta_j = \Delta^{-1}_j \left ( \Delta_j(\beta_j, B), B \right )
\ee
Then the recursive relations for determination of the $\beta_j$ are,
\bea
\beta_{j+1} & = & \min \left ( \max \left ( 
\Delta^{-1}_{j+1} \left ( \alpha_j - \beta_j - 0.03 B, B \right ), 
\sqrt{q} \beta_j \right ), q \beta_j \right )  \nonumber \\
\beta_{j-1} & = & \max \left ( \min \left ( 
\Delta^{-1}_{j-1} \left ( \alpha_j - \beta_j + 0.03 B, B \right ), 
\beta_j/\sqrt{q} \right ), \beta_j/q \right )  \nonumber \\
\beta_0 & = & p
\label{BetajNew}
\eea
where $p$ and $q$ are defined in Eq.\ (\ref{eventemp}).
This completes the AGTO basis set construction for 
the hydrogen iso-electronic sequence in an external constant $B$ field.

To see how faithfully the new prescription reproduces the fully optimized 
exponents, as shown in Fig.\ \ref{ZhuFig1} with empty symbols,
the three solid curves in the figure follow from 
Eqs.\ (\ref{DeltaNew}) through  Eqs.\ (\ref{AlphajNew}).
Indeed, the fits are excellent except in the regions of very tight
functions for orbitals with non-zero angular momentum, where
we adopted the simplified cutoff. Because that is in the weak field
regime, in practice the deviation does not pose any serious problem.
See the discussion above at Eq.\ (\ref{DeltaMin}).
The AGTO basis sets for the three orbitals,  $1s$, 
$2p_{-1}$, and $3d_{-2}$, for the hydrogen atom in $B = 1$ a.u.,
generated according to our construction, Eqs.\ (\ref{DeltaNew}) - (\ref{BetajNew}), 
are represented in the figure  by filled symbols. 
Using those three newly constructed AGTO basis sets, we obtain atomic energies of
-0.8311682 H (Hartree), -0.4565961 H, and  -0.3530471 H, respectively.
Compared to the presumably more accurate algebraic results by
Kravchenko et al.\ \cite{Liberman96PRA}, 
-0.831168896733 H, -0.456597058424 H, and -0.353048025149 H, 
gives residual errors from our AGTO scheme as 
less than 1 $\mu$H. 

More test results are included in Tables \ref{tbl1} and \ref{tbl2}.
We can see that the absolute error in our constructed AGTO basis set
ranges from a fraction of 1 $\mu$H 
to a few $\mu$Hs. While the absolute error increases with increasing 
$B$ field strength, the relative error does not. 
The mean absolute errors (MAEs) of the seven states included in
Table \ref{tbl2} range from 1.4 $\mu$H to 4.2 $\mu$H. 
It is especially noteworthy that not all of those states were used in 
the optimization of exponents nor in the fitting.  
The accuracy of our present basis sets which include only a single sequence 
surpasses that of our previous basis sets, Eqs.\ (\ref{BasSeqs}) through (\ref{b-gamma}),
in which double sequences were included.  Indeed the accuracy of the
new sets is nearly on par with previous basis 
sets that used triple sequences. 

Considering the substantially reduced size of
the new hydrogenic AGTO basis sets, 
we turn in the next subsection to the problem of extending that improvement to 
 many-electron atoms in a strong $B$ field. 

\subsection{\label{ATMBasis} Basis Sets for Many-electron atoms and Ions in $B$ field} 

There are several complications due to the electron-electron interaction in
a many-electron atom or ion. If the $1s$ orbital  
is occupied by one or two electrons, the singular nuclear attractive potential
will be screened, hence be less effective in driving
down the asphericity of tight basis functions in orbitals other than $1s$
than is the case for hydrogen-like ions. This effect dwindles for 
more diffuse basis functions, hence it may be neglected beyond some
point in a sequence of increasingly diffuse exponents. Evidently 
the effect is inversely proportional to the average distance between the orbital
and the nucleus, which is related to the quantum numbers $m$ and $\pi_z$
of the orbital. On the other hand, outer electrons also affect $1s$ electron(s).
Among them, the innermost (tightly bound) non-$1s$ orbital 
($m \ne 0, \pi_z=0$) dominates, so 
it is the only one that we  consider here. 
If the $1s$ orbital is doubly occupied, of course the two $1s$ electrons 
are strongly repulsive to each other, a fact we also shall take into account. 

To describe the aforementioned effects, we generalize 
Eq.\ (\ref{AlphajNew}) to become 
\be
\alpha_{j}  =  \beta_j + f(\beta_j,m_j,\pi_j,B) \Delta_j(\beta_j, B) 
\label{AlphajManye} \; .
\ee
The new feature is the rescaling factor $f$,
\begin{equation}
f(\beta_j,m_j,\pi_j,B) =
\begin{cases}
1 - \frac{N_{1s}}{20}, & \text{if } |m_j| + \pi_j > 0, \text{ and } 
\Delta_j = \alpha_j - \beta_j < \frac{0.14(\pi_j+1.2|m_j|)B}{\pi_j+|m_j|},\\
1 - \frac{1}{20 |m_t|}, & \text{if } |m_j| = \pi_j = 0, m_t \ne 0, \text{and } 
\Delta_j = \alpha_j - \beta_j < 0.17 B,\\
1 - \frac{N_{1s}-1}{20}, & \text{if } |m_j| = \pi_j = 0, N_{1s} > 1, \text{and } 
\Delta_j = \alpha_j - \beta_j < 0.17 B,\\
1,  & \text{otherwise }.
\end{cases}
\label{factorf}
\end{equation}
where $N_{1s} \in $ [0, 2] is the number of electrons that occupy the 
$1s$ orbital, 
including the spin-up and the spin-down electron, and $m_t$ is the 
magnetic quantum number of the occupied innermost tightly bound 
orbital for which the  $z-$parity is even. Observe that Eqs.\ 
(\ref{BetajNew}) remain valid since only the transverse part of
the function is modified.   What Eq.\ (\ref{factorf}) does is
reduce the asphericity slightly at fixed $B$.  That would 
lead to moving the 
points from the upper members of the iso-electronic sequence
slightly downward in the right two-thirds of a many-electron version 
of Fig.\ (\ref{ZhuFig1}). The shift is not enough to warrant a
separate figure.
There is a welcome feature of the $1s$ orbital being occupied. 
Since the nucleus now is screened by the innermost electron cloud, 
the effective repulsive centrifugal potential for non-zero angular 
momentum orbitals
also is smoothed and a cutoff,  Eq.\ (\ref{DeltaMin}), no longer 
is required.

Lastly,  consider the situation in which more than one shell of a single kind
of orbital is occupied. For example, the $1s2s$ configuration for the He atom 
and the ground state $1s^22s$ for the Li atom in a weak $B$ field.
Although we have tried to optimize the transverse exponents by
exploring the $\{ \alpha_j \}$ space while keeping only a single sequence, 
unfortunately we did not succeed in getting adequately accurate 
basis sets. 
The difficulty can be traced to the quite different behaviors of
$1s$ and $2s$ orbitals. It seems unavoidable to have a second
basis function sequence. As we observed above, the severest demands upon 
basis functions are for those having intermediate asphericities.
Bearing this in mind, we choose to limit the second sequence of basis functions 
to have asphericity between $0.03 B$ and $0.225 B$.  Again this is 
a design choice which is based on the fact that the asphericity
curves in Fig.\ \ref{ZhuFig1} are rather flat in the ranges  
$0 \sim 0.03 B$ and $0.225 B \sim 0.25 B$. 
Use of a larger range would result in excessively many basis functions in the
second sequence, hence an un-necessary increase of the basis set size.
For small $B$, the basis functions in the second sequence may collapse 
into (i.e., be approximately linearly dependent with) 
the first sequence. To avoid this problem, a minimum B field
strength of 0.2 a.u.\ is effective 
 as a parameter in calculating $\Delta_j(\beta_j, B)$ for the second sequence.
Thus the second sequence is generated according to
\be
\alpha_{j,2}  =  \beta_j + 0.8 \Delta_j(\beta_j, \max (B, 0.2) ),
\qquad \text{      if } 0.03 B < \Delta_j(\beta_j, B) < 0.225 B.
\label{Alphaj2ndSeq}
\ee
Another technique to avoid basis function collapse is to use $k_j=1$, 
(e.g.,\ $n_{\rho_j} = |m_j| + 2k_j = |m_j| + 2$) for the basis functions,
Eq.\ (\ref{basis}), in the second sequence whenever
$\frac{\alpha_j - \beta_j}{\alpha_j + \beta_j} \le 0.05$. 

Again because of inner electron screening, each electron 
in the atom or the ion feels different effective charges, thus different
reduced $B$ field strength $\gamma$. A detailed analysis can be quite
complicated. Instead, we adopt a rough but effective approximation 
that works quite well. Each electron is assigned an 
effective nuclear charge $Z_{eff}$. For the innermost electrons,
$Z_{eff}$ should be close to the bare nuclear charge, so we simply 
take $Z_{eff} = Z$. 
For other electrons, $Z_{eff}$ is close to the nuclear charge reduced by  
the number of inner-shell electrons. Hence, the reduced field strengths
$\gamma = B/Z_{eff}^2$ are different for each electron.  Those values are used 
for $\gamma$ that appears in Eqs.\ (\ref{DeltaNew}) and (\ref{DeltaNewPara}).
However, notice that the $B$ field experienced by all electrons is the
same.

To here, we have not specified the range of the index $j$ in Eq.\ (\ref{basis}),
which determines the number of the basis functions in a basis set.
The usual computational practice is to use 
a range large enough to ensure that the accuracy
of the basis set is not degraded by insufficient tight 
functions or diffuse functions. More precisely, the incremental error
introduced into the basis set by removing the most diffuse or the most tight
basis function should not exceed the residual error in the unaltered basis set.
This criterion can be met with little difficulty in practice
simply by repeatedly removing
the most diffuse or the most tight basis function and checking 
the calculated results. 
We did tabulate the ranges for different $B$ field strengths
and different orbitals in our computer code, but do not dwell on it
here because the results are both  quite straightforward and not 
very informative. 
One caveat is that in cases for which the electron density
has diffuse, non-zero orbital angular momentum contributions, sometimes 
it is necessary
to extrapolate to negative $j$ values and to include a small number of such 
diffuse basis functions in the basis sets.

We note that many constants which appear in Eqs.\ (\ref{DeltaNew}) to (\ref{Alphaj2ndSeq})
are either chosen by physical plausibility, or were determined by fitting to numerically optimized
data points followed by fine tuning to achieve the best overall performance. 
When the results proved comparatively insensitive to a constant, 
we chose a simpler value
to keep the expressions concise. The forms of these equations are devised
in the hope of encapsulating most of the underlying physics.

We use published HF calculations for atoms in strong $B$ fields to
assess the accuracy of our newly constructed basis sets. The basis
sets errors are given in Tables \ref{tbl3prim} to \ref{tbl7prim} for
atoms He through C and a few ions.  In all cases, the electronic
states are labeled according to their corresponding zero field HF
electronic configurations.  We did not include correlated wavefunction
or DFT calculations because we focus only on basis set construction in
this study, and calculations including electron correlation would
complicate our comparisons.
However, we note that
the basis sets constructed according to the new procedure
are intended for use both in correlated 
wavefunction and DFT calculations, not just  HF calculations.

There is an issue of reference data.  For the He atom, Ruder and
coworkers did extensive HF calculations and gave abundant tabulated
data\cite{Ruder94book}. Jones, Ortiz, and Ceperley also used AGTO
basis functions but different basis sets \cite{Ceperley96PRA}.  Zhao
and Stancil used a B-spline basis and quadruple-precision calculations
to obtain very accurate HF energies for $0 \le B \le 100$ a.u., but
their data is only available for the $1s^2$ state
\cite{ZhaSta-PRA-2008}.  Using different sources as reference data
will introduce non-uniformity of accuracies, thus jeopardize clarity of 
comparison. Hence, we decided to use our own recent results as
reference.  They were obtained by using the multiple sequence AGTO
basis functions \cite{ZhuZhaTri-PRA-2014}. They were in good agreement
with other published results, and even have higher accuracy for
some states (lower HF energies compared to \cite{Ceperley96PRA}).

Basis set errors for the He atom in ${\bf B}$ fields are shown in
Table \ref{tbl3prim}.  Absolute errors range from a few to a couple of
hundreds of $\mu$H, with an average of a few tens of
$\mu$H. Understandably they are larger than the basis set errors for H
atom, but the relative errors are not.
The MAE for the $1s2s$ state seems significantly smaller than 
those of other states, 
simply because double sequences were employed for that case alone. At low 
field strength $B = 0.1 a.u.$, 
there are several negative values in the table, which means that our current
single sequence basis sets give lower HF energies than 
the reference calculations that used 
the multiple sequence basis sets of Eq.\ (\ref{BasSeqs}). The improvement is obvious.
The test on this simplest two-electron system shows that our constructed basis
sets work as well as for a single electron system, H atom.

Tables \ref{tbl4prim} to \ref{tbl7prim} give the basis set errors 
for the Li, Be, B, C atoms and the positive ions, Li$^+$, Be$^+$, and B$^+$, 
in field strengths $0 \le B \le 2000$ a.u. Comparison data for atoms and ions 
with $ Z \ge 3 $ are from the series studies by Ivanov and Schmelcher, 
who used two-dimensional (2D) mesh methods \cite{Ivanov}.  
Exceptions are the 1s$^2$2s$^2$ state of the Be atom and  B$^+$ ion, 
and the 1s$^2$2s$^2$2p$_{-1}$ state of the B atom, for which they allowed 
asymmetrical wavefunctions for 2s$^2$ electrons with respect to 
the $z = 0$ plane \cite{Ivanov}, whereas we required definite $z$ parity, $\pi_{z} = 0,1$. 
For those three instances, the references are take from our own calculations by using
very large basis sets including multiple sequences \cite{ZhuZhaTri-PRA-2014}.  
We also supplemented the reference data points for the Li atom at $B = 2000$ a.u.

Basis set errors in these calculations range from a few hundredths to
a few  mH. As mentioned before, negative values 
signify that we obtained lower HF energies with our new basis sets
than those from our chosen references. Although absolute errors vary a lot,
relative errors are not very different from those for the H and He
atoms, and do not change much with increasing $B$ field strength
(except when the total energy of the atom or ion happens to be close
to zero), or for different atoms.  This is evidence for the
uniformity of the quality of our newly constructed basis sets for
different atoms in different $B$ field strengths, a welcome feature.
Inspection of data in Tables \ref{tbl4prim} to
\ref{tbl7prim} shows a few irregularities of error distribution. For
example, in the Li atom the error for the $1s2s2p_{-1}$ state at
$B=500$ a.u. is 3.14 mH, much larger than the errors at $B=200$ a.u. (0.33
mH) and at $B=1000$ a.u. (0.92 mH). We surmise that the reason is that our
chosen reference data have non-uniform accuracies.  Using our own
results from multiple sequence calculations
\cite{ZhuZhaTri-PRA-2014}, the basis set errors are much more
uniform: 
0.24mH, 0.30mH, 0.28mH, for $B = 200, 500, 1000 $ a.u., respectively 
(not included in the Table).  Another example is the
$1s2p_02p_{-1}$ state at $B=200$ a.u., for which we got a total HF
energy 1.25 mH lower than the reference result.  
The MAEs for each atom also vary
considerably, but the relative errors for each atom are always a few to 25
parts in a million, again suggesting 
that our new basis set construction provides relatively uniform accuracy for different atoms.

\section{\label{Concl}Summary and Conclusions}

AGTO basis functions are well adapted to the treatment of atoms and molecules in
arbitrarily strong magnetic field \cite{Schmelcher88PRA}. 
But the complexity and high cardinality of existing AGTO sets has 
impeded their use in such calculations. 
This criticism is valid for our earlier scheme \cite{ZhuZhaTri-PRA-2014}, 
even though it provided the substantial advantage of a fixed, prescribed
basis set rather than case-by-case optimization.  The present work
represents a significant technical advance based upon  
detailed physical analysis and extensive numerical exploration. 
The combination enables establishment of a few key principles to guide 
basis set construction.  Those principles are: (1) The
more diffuse the function (smaller exponents), the larger the
asphericity (relative difference between the transverse and
longitudinal exponents of the basis function), with the limit of $B/4$
for the most diffuse functions, which is actually a Landau orbital;
(2) More densely spaced functions are put in the range of
exponents across which
asphericities change most rapidly; (3) When more than one
atomic shell is occupied by electrons for a single kind of orbital, a second
sequence of basis functions is required, but the second sequence can
be limited to the range over which asphericities change rapidly, e.g., $[0.03B,
  0.225B]$; (4) The interaction of a $1s$ electron with other electrons,
or the interaction between two electrons when the $1s$
orbital is doubly occupied, can slightly reduce the asphericity of the
tight basis functions in the basis set.

With these guidelines, we first investigated, in detail, 
the hydrogen iso-electronic sequence in an arbitrary $B$ field. Exploration
of the transverse exponent  parameter space $\{ \alpha_j \}$, 
with full numerical optimization leads to fitting the results
with a newly designed expression which takes into account the
orbital quantum numbers. By fine tuning  the constants in the expression,
we obtained Eqs.\ (\ref{DeltaNew}) - (\ref{BetajNew}).
The residual basis error from this newly constructed AGTO basis sets
is no more than a few $\mu$Hs, while the relative error is only a few millionths.

For light multiple-electron atoms and ions, we further considered the strong
electron-electron repulsion at the vicinity of nucleus, the
aforementioned point (4), and slightly modified the basis sets of
single electron systems, leading to Eqs.\ (\ref{AlphajManye}) and
(\ref{factorf}), which depend on orbital quantum numbers and
electron occupation numbers, as well as on the $B$ field strength and
nuclear charge $Z$.  If more than one shell of a kind of orbital is
occupied, we find it necessary to supplement the main sequence with 
a second basis function sequence in the exponent range of rapidly changing asphericity
of primitives, Eqs.\ (\ref{Alphaj2ndSeq}).  While absolute basis set
errors vary noticeably, from a few hundredths to a few of mH, the relative
errors are similar for different atoms and ions in a wide range of $B$
field strengths, indicating that our currently constructed basis sets
have satisfactorily near-uniform accuracy.

We have not studied contraction of the newly constructed basis
primitives, but believe it could be done without any significant 
difficulty.  Contracting primitives can  reduce the number of
basis functions, a desirable step for correlated wavefunction
calculations and one which can be helpful in DFT calculations.
Without contraction,  the residual basis set errors 
in the newly constructed
basis sets are two to three orders of magnitude smaller than the
electron correlation energies in the atomic and ionic systems
investigated here. Hence our basis sets should be sufficiently
accurate for correlated wavefunction or DFT calculations.

For the convenience of our readers, the supplemental
material \cite{Supp} provides (under GPL) our computer code (written in C) 
which generated all the AGTO basis sets used in this work. It is also
available by download from www.qtp.ufl.edu/ofdft.

\begin{acknowledgments}
Helpful discussions with A.\ Savin (CNRS), Xin Xu (Fudan University),
X.-Y. Pan (Ningbo University), and Jian Wang (Huzhou University) 
are acknowledged with thanks. 
This work was funded by Zhejiang Provincial Natural Science Foundation 
of China under grant No.\ LY13A050002 (W.\ Zhu),
in part by the National Natural Science Foundation of China grant No.\ 
11474081 (W.\ Zhu),
and in part by U.S.\ Dept.\ of Energy grant DE-SC-0002139 (S.B.\ Trickey).
\end{acknowledgments}

\newpage

\scriptsize

\begin{table}
\caption{\label{tbl1}Comparison of the H atomic energies in $\mathbf B$ fields between our present study
using a single sequence of AGTO basis functions and the more accurate algebraic results as reference values.
(Atomic energies in Hartree, absolute errors in $\mu$H, relative errors in millionth, and $\mathbf B$ in a.u.)}
{
\scriptsize
\linespread{0.2}
\begin{tabular}{llllrrrrrrrr} \hline  \hline
  &        &\multicolumn{4}{c}{H (1s)} &  & \multicolumn{4}{c}{H (2p$_{0}$)}\\
\cline{3-6} \cline{8-11}
  & $B$(a.u.) & present & ref.$^a$\tnote{a}  &  error($\mu$H)  & relative error & \hspace{2em} & present & ref.$^a$\tnote{a}  &  error($\mu$H)  & relative error  \\ \cline{3-6} \cline{8-11}
  &    0     &  -0.49999993  &  -0.50000000  &  0.07  & 0.14 &   &  -0.12499988  &  -0.12500000  &  0.12  &    1.0\\
  &    0.01  &  -0.50497493  &  -0.50497500  &  0.07  & 0.14 &	 &  -0.12985022  &  -0.12985042  &  0.20  &    1.5\\
  &    0.02  &  -0.50989998  &  -0.50990004  &  0.06  & 0.14 &	 &  -0.13440604  &  -0.13440647  &  0.43  &    3.2\\
  &    0.05  &  -0.52437664  &  -0.52437671  &  0.07  & 0.13 &	 &  -0.14646326  &  -0.14646484  &  1.6  &   10.8\\
  &    0.1   &  -0.54752639  &  -0.54752648  &  0.09  & 0.17 &	 &  -0.16240820  &  -0.16241008  &  1.9  &   11.5\\
  &    0.2   &  -0.59038138  &  -0.59038157  &  0.19  & 0.31 &	 &  -0.18518248  &  -0.18518404  &  1.6  &    8.5\\
  &    0.5   &  -0.69721007  &  -0.69721054  &  0.47  &  0.7 &	 &  -0.2247593  &  -0.22476034  &  1.0  &    4.5\\
  &    1     &  -0.83116821  &  -0.83116890  &  0.69  &  0.8 &	 &  -0.2600041  &  -0.26000662  &  2.5  &    9.7\\
  &    2     &  -1.0222124  &   -1.02221391  &  1.5  &  1.5 &	 &  -0.2977091  &  -0.29771097  &  1.9  &    6.2\\
  &    5     &  -1.3803975  &   -1.38039887  &  1.4  &  1.0 &	 &  -0.3476164  &  -0.34761778  &  1.3  &    3.8\\
  &   10     &  -1.7477962  &   -1.74779716  &  0.9  &  0.5 &	 &  -0.3826474  &  -0.38264985  &  2.4  &    6.4\\
  &   20     &  -2.2153967  &   -2.21539852  &  1.8  &  0.8 &	 &  -0.4133743  &  -0.41337773  &  3.4  &    8.3\\
  &   50     &  -3.0178572  &   -3.01786071  &  3.6  &  1.2 &	 &  -0.4456826  &  -0.44568511  &  2.5  &    5.6\\
  &  100     &  -3.7897977  &   -3.78980424  &  6.5  &  1.7 &	 &  -0.4636165  &  -0.46361776  &  1.3  &    2.8\\
  &  200     &  -4.7271409  &   -4.72714511  &  4.2  &  0.9 &	 &  -0.4765313  &  -0.47653200  &  0.7  &    1.5\\
  &  500     &  -6.2570816  &   -6.25708767  &  6.0  &  1.0 &	 &  -0.4875069  &  -0.48750710  &  0.2  &    0.5\\
  & 1000     &  -7.6624154  &   -7.66242325  &  7.8  &  1.0 &	 &  -0.4924948  &  -0.49249500  &  0.2  &    0.4\\
  & 2000     &  -9.3047557  &  -9.30476508   &  9.4  &  1.0 & \\
  & 4000     & -11.2041340  & -11.20414521   & 11.2  &  1.0 & \\ \hline
  &        &\multicolumn{4}{c}{H (2p$_{-1}$)} &  & \multicolumn{4}{c}{H (3d$_{-2}$)}\\
\cline{3-6} \cline{8-11}
  & $B$(a.u.) & present & ref.$^a$\tnote{a}  &  error($\mu$H)  & relative error & \hspace{2em} & present & ref.$^a$\tnote{a}  &  error($\mu$H)  & relative error  \\ \cline{3-6} \cline{8-11}
  &    0    & -0.12499988 &  -0.12500000  & 0.12 &  1.0 &   &    -0.05555511 &  -0.05555556 &   0.45 &   8.0\\
  &    0.01 & -0.13470101 &  -0.13470114  & 0.13 &  1.0 &   &    -0.06924669 &  -0.06924718 &   0.49 &   7.2\\
  &    0.02 & -0.14381741 &  -0.14381761  & 0.20 &  1.4 &   &    -0.08068515 &  -0.08068587 &   0.72 &   8.9\\
  &    0.05 & -0.16805783 &  -0.16805819  & 0.36 &  2.2 &   &    -0.10688868 &  -0.10688875 &   0.07 &   0.7\\
  &    0.1  & -0.20084544 &  -0.20084567  & 0.23 &  1.2 &   &    -0.13783895 &  -0.13783952 &   0.57 &   4.1\\
  &    0.2  & -0.25053886 &  -0.25053910  & 0.24 &  0.9 &   &    -0.18132001 &  -0.18132061 &   0.60 &   3.3\\
  &    0.5  & -0.34947668 &  -0.34947730  & 0.62 &  1.8 &   &    -0.26438927 &  -0.26438955 &   0.28 &   1.1\\
  &    1    & -0.45659614 &  -0.45659706  & 0.92 &  2.0 &   &    -0.35304715 &  -0.35304803 &   0.88 &   2.5\\
  &    2    & -0.59961193 &  -0.59961277  & 0.84 &  1.4 &   &    -0.4711704 &  -0.47117193 &   1.5 &   3.2\\
  &    5    & -0.85983170 &  -0.85983262  & 0.92 &  1.1 &   &    -0.6867994 &  -0.68680252 &   3.1 &   4.5\\
  &   10    & -1.1254206  &  -1.12542234  & 1.7  &  1.5 &   &    -0.9082117 &  -0.90821478 &   3.1 &   3.4\\
  &   20    & -1.4655037  &  -1.46550855  & 4.8  &  3.3 &   &    -1.1936274 &  -1.19363318 &   5.8 &   4.9\\
  &   50    & -2.0568422  &  -2.05684667  & 4.5  &  2.2 &   &    -1.6943146 &  -1.69432125 &   6.7 &   3.9\\
  &  100    & -2.6347533  &  -2.63476067  & 7.4  &  2.8 &   &    -2.1881638 &  -2.18816724 &   3.5 &   1.6\\
  &  200    & -3.3471360  &  -3.34714523  & 9.3  &  2.8 &   &    -2.8019958 &  -2.80200003 &   4.3 &   1.5\\
  &  500    & -4.5312368  &  -4.53124638  & 9.6  &  2.1 &   &    -3.8323855 &  -3.83239006 &   4.5 &   1.2\\
  & 1000    & -5.6384097  &  -5.63842108  &11.4  &  2.0 &   &    -4.8051057 &  -4.80511067 &   5.0 &   1.0\\ \hline \hline
\end{tabular}}
    \begin{tablenotes}
\tiny
       \item[a]$^a$ Data are from Ref.\ \cite{Liberman96PRA}.
    \end{tablenotes}
\end{table}
\newpage

\begin{table}
\caption{\label{tbl2} Basis set errors for the H atom in ${\bf B}$ fields$^a$.  (absolute errors in $\mu$H, 
numbers in parenthesis are the relative errors in millionths, and $\mathbf B$ in a.u.)}
\begin{threeparttable}
{\scriptsize
\begin{tabular}{llrrrrrrrrr} \hline  \hline
  & $B$(a.u.) & 1s &   2p$_{0}$   &   2p$_{-1}$  &  3d$_{-1}$  &   3d$_{-2}$  &  4f$_{-3}$   &   5g$_{-4}$\\  \hline
 & 0   & 0.1( 0.1) &   0.1( 1.0) &   0.1( 1.0) &   0.4( 8.0) &  0.4( 8.0)  &  0.7(  24)  &  2.0(  99)  \\ 
 & 0.01& 0.1( 0.1) &   0.2( 1.5) &   0.1( 1.0) &   0.9(  14) &  0.5( 7.2)  &  1.1(  23)  &  1.7(  45)  \\ 
 & 0.02& 0.1( 0.1) &   0.4( 3.2) &   0.2( 1.4) &   1.7(  23) &  0.7( 8.9)  &  0.5( 7.9)  &  0.9(  19)  \\ 
 &0.05 & 0.1( 0.1) &   1.6(  11) &   0.4( 2.2) &   0.4( 4.8) &  0.1( 0.7)  &  0.2( 2.2)  & -0.1( 1.8)  \\
 & 0.1 & 0.1( 0.2) &   1.9(  12) &   0.2( 1.2) &   0.2( 1.8) &  0.6( 4.1)  &  1.1( 9.9)  &  0.4( 3.9)  \\
 & 0.2 & 0.2( 0.3) &   1.6( 8.5) &   0.2( 0.9) &   0.2( 1.9) &  0.6( 3.3)  &  0.7( 4.6)  &  1.0( 7.7)  \\
 & 0.5 & 0.5( 0.7) &   1.0( 4.5) &   0.6( 1.8) &   2.1(  12) &  0.3( 1.1)  &  1.1( 4.7)  &  6.3(  32)  \\
 &   1 & 0.7( 0.8) &   2.5( 9.7) &   0.9( 2.0) &   3.2(  15) &  0.9( 2.5)  &  2.3( 7.8)  &  7.7(  29)  \\
 &   2 & 1.5( 1.5) &   1.9( 6.2) &   0.8( 1.4) &   2.5(  10) &  1.5( 3.2)  &  3.3( 8.3)  &  8.0(  22)  \\
 &   5 & 1.4( 1.0) &   1.3( 3.8) &   0.9( 1.1) &   6.7(  23) &  3.1( 4.5)  &  4.7( 7.9)  &  4.8( 9.1)  \\
 &  10 & 0.9( 0.5) &   2.4( 6.4) &   1.7( 1.5) &   6.8(  20) &  3.1( 3.4)  &  4.0( 5.0)  &  2.5( 3.5)  \\
 &  20 & 1.8( 0.8) &   3.4( 8.3) &   4.8( 3.3) &   7.6(  20) &  5.8( 4.9)  &  7.2( 6.9)  &  3.0( 3.2)  \\
 &  50 & 3.6( 1.2) &   2.5( 5.6) &   4.5( 2.2) &   4.7(  11) &  6.7( 3.9)  &  7.3( 4.9)  &  4.2( 3.1)  \\
 & 100 & 6.5( 1.7) &   1.3( 2.8) &   7.4( 2.8) &   2.9( 6.4) &  3.5( 1.6)  &  4.5( 2.3)  &  9.2( 5.3)  \\
 & 200 & 4.2( 0.9) &   0.7( 1.5) &   9.3( 2.8) &   1.5( 3.2) &  4.3( 1.5)  &  5.3( 2.1)  &  5.9( 2.6)  \\
 & 500 & 6.0( 1.0) &   0.2( 0.5) &   9.6( 2.1) &   0.5( 1.1) &  4.5( 1.2)  &  6.1( 1.8)  &  7.1( 2.3)  \\
 &1000 & 7.8( 1.0) &   0.2( 0.4) &  11.4( 2.0) &   0.1( 0.2) &  5.0( 1.0)  &  6.4( 1.5)  &  7.4( 1.9)  \\ \hline
 & MAE & 2.09(0.7) &  1.36( 5.1) &  3.12( 1.8) &  2.49(  10) &  2.45( 3.6) &  3.32( 7.3) &  4.24(  17)  \\ \hline \hline
\end{tabular}}
 \begin{tablenotes}
\scriptsize
  \item[a] Reference values used to deduce basis set errors are from Ref.\ \cite{Liberman96PRA}
for the $1s$, $2p_{0}$, $2p_{-1}$, $3d_{-1}$ and $3d_{-2}$ states. For the states
$4f_{-3}$ and $5g_{-4}$, reference data are taken from our calculation with very large basis sets 
which include multiple sequences (through five) of Eq.\ (\ref{BasSeqs}). 
There is slight
improvement over our earlier published results in the low-$B$ field region \cite{ZhuZhaTri-PRA-2014}. 
 \end{tablenotes}
\end{threeparttable}
\end{table}

\begin{table}
\caption{\label{tbl3prim} Basis set errors for the He atom in ${\bf B}$ fields$^a$.  (absolute errors in $\mu$H, 
numbers in parenthesis are the relative errors in millionths, and $\mathbf B$ in a.u.)}
\begin{threeparttable}
{\scriptsize
\begin{tabular}{llrrrrrrrrr} \hline  \hline
 & $B$(a.u.) &  1s$^2$& 1s2s&  1s2p$_{0}$& 1s2p$_{-1}$&  1s3d$_{-1}$&  1s3d$_{-2}$&   1s4f$_{-2}$&   1s4f$_{-3}$&   1s5g$_{-3}$\\  \hline

 &   0 &   -1(0.3) &   1(0.4) &  20(9.3) &   5(2.4) &   0(0.2) &   0(0.2) &   1(0.5) &   0(0.2) &  31( 15)\\ 
 & 0.1 &    0(0.0) &   0(0.0) &  25( 11) &   8(3.7) &  -5(2.5) &   7(3.4) &  -8(3.9) & -13(5.9) & -27( 13)\\ 
 & 0.2 &    0(0.1) &   4(1.8) &  39( 17) &  19(8.1) &  10(4.7) &  18(7.9) &   7(3.0) &   7(3.2) &   5(2.1)\\ 
 & 0.5 &    1(0.4) &  13(5.2) &  53( 21) &  49( 19) &  11(4.6) &  12(4.8) &   7(2.8) &   8(3.4) &  20(8.5)\\ 
 &   1 &    2(0.6) &  11(4.2) &  34( 13) &  61( 21) &  17(6.6) &  18(6.4) &  18(7.0) &  13(4.9) &  46( 18)\\ 
 &   2 &    2(1.1) &   9(3.0) &  27(8.6) &  62( 18) &  19(6.3) &  16(4.9) &   9(2.9) &  19(6.0) &   6(2.0)\\ 
 &   5 &    2(4.4) &   7(1.8) &  11(2.9) &  45(9.7) &  19(5.0) &  17(3.9) &   7(1.8) &  27(6.6) &   1(0.2)\\ 
 &  10 &   20(6.6) &   4(0.8) &   8(1.7) &  29(4.9) &  17(3.5) &  29(5.5) &   8(1.7) &  53( 10) &   3(0.6)\\ 
 &  20 &   17(1.5) &   0(0.1) &   4(0.6) &  38(5.1) &  10(1.6) &  30(4.4) &   4(0.7) &  40(6.0) &   3(0.5)\\ 
 &  50 &   27(0.7) &  -3(0.4) &  11(1.3) &  65(6.3) &  19(2.4) &  35(3.7) &   9(1.2) &  33(3.6) &   7(0.8)\\ 
 & 100 &   43(0.5) &   0(0.0) &  16(1.6) & 101(7.7) &  20(2.0) &  48(4.0) &  14(1.4) &  36(3.1) &  12(1.2)\\ 
 & 200 &   72(0.4) &  -6(0.4) &  13(1.1) & 133(8.0) &  16(1.3) &  67(4.3) &  12(1.0) &  46(3.1) &  11(0.9)\\ 
 & 500 &  123(0.3) &  -4(0.3) &  23(1.4) & 178(7.9) &  11(0.7) &  89(4.3) &  10(0.6) &  62(3.1) &  10(0.6)\\ 
 &1000 &  179(0.2) &  -1(0.1) &  22(1.1) & 204(7.3) &  18(0.9) & 105(4.0) &  17(0.8) &  74(3.0) &  18(0.9)\\ 
 &2000 &  189(0.1) &   1(0.1) &  25(1.0) & 220(6.3) &  25(1.0) & 111(3.4) &  24(1.0) &  85(2.7) &  24(0.9)\\ \hline  
 & MAE &   45(1.1) &   4(1.2) &  22(6.2) &  81(9.0) &  14(2.9) &  40(4.3) &  10(2.0) &  34(4.3) &  15(4.3)\\  \hline \hline 
\end{tabular}}
 \begin{tablenotes}
\scriptsize
  \item[a] Reference values used to deduce basis set errors are from Ref.\ \cite{ZhuZhaTri-PRA-2014},
in which very large basis sets including multiple sequences of Eq.\ (\ref{BasSeqs}) were used. Recalculation slightly improved over our earlier published 
results in the low $B$-field region, $0 \le B \le 1 a.u.$ 
 \end{tablenotes}
\end{threeparttable}
\end{table}

\begin{table}
\caption{\label{tbl4prim} Basis set errors for the  Li singly positive ion and Li atom in ${\bf B}$ fields$^a$.  (absolute errors in mH, numbers in parenthesis are the relative errors in millionth, and $\mathbf B$ in a.u.)}
\begin{threeparttable}
{\scriptsize
\begin{tabular}{llrrrrrrrrr} \hline  \hline
&           &\multicolumn{2}{c}{Li$^+$} &\hspace{1em} &\multicolumn{5}{c}{Li}\\ \cline{3-4} \cline{6-10}
& $B$(a.u.) & 1s$^2$ & \hspace{2em}1s2p$_{-1}$ &\hspace{1em} & \hspace{1em}1s$^2$2s  &1s$^2$2p$_{-1}$ &1s2s2p$_{-1}$ \hspace{1em} & 1s2p$_0$2p$_{-1}$ &1s2p$_{-1}$3d$_{-2}$  \\  \hline
&   0 &  0.01( 0.8) &   0.02( 4.3) & &   0.00( 0.1) &   0.01( 1.7) &  0.03( 4.7) &   0.01( 2.4) &  0.04( 6.9) \\
& 0.1 & -0.00( 0.3) &   0.03( 5.8) & &   0.01( 1.4) &   0.02( 3.2) &  0.04( 7.4) &   0.02( 3.7) &  0.07(  13) \\
& 0.2 &  0.00( 0.3) &   0.03( 5.9) & &   0.00( 0.6) &   0.03( 4.5) &  0.11(  20) &   0.02( 3.8) &  0.10(  18) \\
& 0.5 &  0.00( 0.3) &   0.04( 6.6) & &   0.01( 1.7) &   0.05( 6.2) &  0.03( 5.1) &   0.06(  10) &  0.07(  11) \\
&   1 &  0.00( 0.4) &   0.06(  10) & &   0.02( 2.1) &   0.09(  11) &  0.05( 8.3) &   0.15(  23) &  0.09(  14) \\
&   2 &  0.01( 1.1) &   0.07(  10) & &   0.01( 1.1) &   0.10(  13) &  0.11(  15) &   0.20(  27) &  0.10(  14) \\
&   5 &  0.02( 2.7) &   0.07( 7.7) & &   0.01( 2.3) &   0.11(  16) &  0.01( 1.5) &   0.19(  21) &  0.11(  12) \\
&  10 &  0.11(  36) &   0.03( 2.9) & &  -0.07(  21) &   0.14(  30) &  0.05( 4.2) &   0.15(  14) &  0.10( 8.5) \\
&  20 &  0.10(  25) &   0.04( 3.0) & &  -0.00( 0.4) &   0.13(  74) &  0.08( 6.1) &   0.09( 6.1) &  0.07( 4.6) \\
&  50 &  0.10( 3.6) &   0.13( 7.1) & &  -0.26( 9.3) &   0.15( 6.1) &  0.13( 7.0) &   0.15( 7.7) &  0.08( 4.0) \\
& 100 &  0.04( 0.5) &   0.14( 6.0) & &   0.65( 9.1) &   0.21( 3.1) &  0.85(  35) &   0.01( 0.4) &  0.14( 5.4) \\
& 200 &  0.05( 0.3) &   0.20( 6.6) & &  -0.75( 4.5) &   0.25( 1.6) &  0.33(  11) &  -1.25(  40) &  0.34( 9.7) \\
& 500 &  0.09( 0.2) &   0.32( 7.7) & &  -1.96( 4.3) &   0.39( 0.9) &  3.14(  75) &   0.83(  20) &  0.52(  11) \\
&1000 &  0.13( 0.1) &   0.33( 6.2) & &  12.32(  13) &   0.74( 0.8) &  0.92(  18) &   0.93(  18) &  0.61(  10) \\
&2000 &  0.23( 0.1) &   0.36( 5.6) & &   0.00( 0.0) &   0.94( 0.5) &  0.33( 5.0) &   0.43( 6.5) &  0.66( 8.8) \\  \hline
& MAE &  0.06( 4.8) &   0.13( 6.4) & &   1.07( 4.7) &   0.22(  12) &  0.41(  15) &   0.30(  14) &  0.21(  10) \\  \hline \hline
\end{tabular}}
 \begin{tablenotes}
\scriptsize
  \item[a] Reference values used to deduce basis set errors are from Ref.\ \cite{Ivanov}.
For $B = 2000$ a.u., reference data are taken from our calculation with very large basis sets 
which include multiple sequences of Eq.\ (\ref{BasSeqs}) \cite{ZhuZhaTri-PRA-2014}. 
 \end{tablenotes}
\end{threeparttable}
\end{table}

\begin{table}
\caption{\label{tbl5prim} Basis set errors for the  Be singly positive ion and Be atom in ${\bf B}$ fields$^a$.  (Absolute errors in mH, numbers in parenthesis are the relative errors in millionths, and $\mathbf B$ in a.u.)}
\begin{threeparttable}
{\scriptsize
\begin{tabular}{llrrrrrrrrr} \hline  \hline
&           &\multicolumn{3}{c}{Be$^+$} &\hspace{2em} &\multicolumn{4}{c}{Be}\\ \cline{3-5} \cline{7-10}
& $B$(a.u.) & 1s$^2$2s & 1s$^2$2p$_{-1}$ & 1s2p$_{-1}$3d$_{-2}$ & &1s$^2$2s$^2$&\hspace{2em}1s$^2$2s2p$_{-1}$&1s$^2$2p$_{-1}$3d$_{-2}$&1s2p$_{-1}$3d$_{-2}$4f$_{-3}$\\  \hline
&   0 &   0.01( 0.6) &   0.02( 1.1) &   0.05( 5.8) & &  -0.00( 0.1) &   0.03( 1.7) &   0.04( 2.8) &   0.08( 8.7)  \\
& 0.5 &   0.02( 1.3) &   0.05( 3.1) &   0.11(  10) & &   0.01( 1.0) &   0.03( 2.1) &   0.07( 4.7) &   0.08( 7.5)  \\
&   1 &   0.03( 2.2) &   0.09( 6.1) &   0.09( 7.7) & &   0.05( 3.4) &   0.05( 3.3) &   0.11( 7.1) &   0.10( 8.2)  \\
&   2 &   0.05( 3.7) &   0.13( 8.9) &   0.10( 7.9) & &   0.07( 5.2) &   0.06( 4.0) &   0.18(  12) &   0.11( 8.4)  \\
&   5 &   0.05( 3.4) &   0.18(  12) &   0.11( 7.2) & &   0.09(  10) &   0.07( 4.7) &   0.25(  15) &   0.18(  11)  \\
&  10 &   0.03( 2.8) &   0.16(  12) &   0.12( 6.5) & &   0.07(  44) &   0.12( 8.7) &   0.27(  18) &   0.14( 6.8)  \\
&  20 &   0.04( 7.0) &   0.22(  24) &   0.11( 4.7) & &   0.07( 5.2) &   0.12(  12) &   0.33(  30) &   0.16( 6.2)  \\
&  50 &  -0.06( 3.6) &   0.34(  32) &   0.24( 7.2) & &   0.06( 0.9) &   0.30(  29) &   0.39(  50) &   0.29( 8.3)  \\
& 100 &  -0.05( 0.9) &   0.25( 5.1) &   0.30( 7.1) & &   0.07( 0.4) &   0.57(  12) &   0.40( 8.6) &   0.40( 8.9)  \\
& 200 &   0.15( 1.1) &   0.31( 2.3) &   0.41( 7.5) & &  -0.07( 0.2) &  -0.60( 4.4) &   0.54( 4.1) &   0.62(  11)  \\
& 500 &  -2.91( 6.8) &   0.56( 1.4) &   0.66( 8.8) & &  -0.41( 0.4) &   1.23( 3.0) &   1.19( 2.9) &   0.94(  12)  \\
&1000 &  12.09(  13) &   0.74( 0.8) &   0.83( 8.7) & &  -0.92( 0.5) &  -1.88( 2.1) &   0.60( 0.7) &   1.22(  12)  \\
&2000 & -18.31( 9.7) &   0.95( 0.5) &   0.93( 7.7) & &  -3.09( 0.8) & -30.10(  16) &  -2.53( 1.4) &   1.30(  10)  \\ \hline 
& MAE &   2.60( 4.3) &   0.31( 8.4) &   0.31( 7.5) & &   0.38( 5.6) &   2.70( 8.0) &   0.53(  12) &   0.43( 9.1)  \\ \hline  \hline 
\end{tabular}}
\begin{tablenotes}
\scriptsize
  \item[a] Reference values used to deduce basis set errors are from Ref.\ \cite{Ivanov}.
For the 1s$^2$2s$^2$ state of Be atom, reference values are from our own calculation by using
very large basis sets including multiple sequences \cite{ZhuZhaTri-PRA-2014}. See text for details.
 \end{tablenotes}
\end{threeparttable}
\end{table}

\begin{table}
\caption{\label{tbl6prim} Basis set errors for the  B singly positive ion and B atom in ${\bf B}$ fields$^a$.  (Absolute errors in mH, numbers in parenthesis are the relative errors in millionths, and $\mathbf B$ in a.u.)}
\begin{threeparttable}
{\scriptsize
\begin{tabular}{llrrrrrrrrrrrr} \hline  \hline
&           &\multicolumn{4}{c}{B$^+$} &\hspace{2em} &\multicolumn{6}{c}{B}\\ \cline{3-6} \cline{8-13}
&      &        & 1s$^2$ & 1s$^2$ & 1s2p$_{-1}$  &\hspace{1em}&1s$^2$2s$^2$ &1s$^2$2s &1s$^2$2s &1s$^2$2p$_0$ &1s$^2$2p$_{-1}$ &1s$^2$2p$_{-1}$ \\
& $B$(a.u.)& 1s$^2$2s$^2$ & 2s2p$_{-1}$ &2p$_{-1}$3d$_{-2}$ &3d$_{-2}$4f$_{-3}$& &2p$_{-1}$&2p$_0$2p$_{-1}$&2p$_{-1}$3d$_{-2}$&2p$_{-1}$3d$_{-2}$&3d$_{-2}$4f$_{-3}$&3d$_{-2}$4f$_{-3}$5f$_{-4}$ \\ \hline
& 0.0 &  0.00(0.0)& 0.15(6.4)& 0.06(2.8)& 0.10(6.7)& & 0.11(4.5)& 0.10(4.0)& 0.18(7.4)& 0.07(2.7)& 0.14(6.0)& 0.14(9.1)\\
& 0.1 & -0.00(0.0)& 0.25( 10)& 0.17(7.2)& 0.29( 19)& & 0.20(8.3)& 0.17(7.1)& 0.43( 18)& 0.23(9.4)& 0.38( 16)& 0.21( 13)\\
& 0.2 & -0.00(0.2)& 0.37( 15)& 0.23(9.6)& 0.20( 12)& & 0.49( 20)& 0.33( 13)& 0.40( 16)& 0.20(8.1)& 0.23(9.6)& 0.19( 12)\\
& 0.5 &  0.01(0.4)& 0.11(4.6)& 0.10(4.2)& 0.17( 10)& & 0.17(7.0)& 0.22(8.9)& 0.13(5.2)& 0.19(7.5)& 0.15(6.2)& 0.14(8.1)\\
&   1 &  0.04(1.8)& 0.05(2.1)& 0.15(6.1)& 0.14(7.8)& & 0.12(5.0)& 0.18(7.1)& 0.08(3.3)& 0.22(8.8)& 0.17(6.8)& 0.13(7.0)\\
&   2 &  0.10(4.4)& 0.07(2.7)& 0.19(7.3)& 0.13(6.6)& & 0.25( 10)& 0.18(6.8)& 0.10(3.7)& 0.23(8.8)& 0.21(7.9)& 0.14(7.1)\\
&   5 &  0.19( 10)& 0.13(5.1)& 0.34( 13)& 0.15(6.1)& & 0.23( 11)& 0.24(9.1)& 0.16(5.8)& 0.26(9.4)& 0.35( 13)& 0.29( 12)\\
&  10 &  0.11(8.7)& 0.13(4.9)& 0.41( 15)& 0.18(6.1)& & 0.21( 13)& 0.21(7.9)& 0.28( 10)& 0.22(8.0)& 0.47( 17)& 0.21(7.2)\\
&  20 &  0.10( 47)& 0.19(8.5)& 0.56( 22)& 0.20(5.5)& & 0.37(135)& 0.21(8.8)& 0.33( 13)& 0.33( 13)& 0.68( 26)& 0.29(7.8)\\
&  50 &  0.20(4.0)& 0.51( 77)& 0.86( 81)& 0.46(9.3)& & 0.89( 21)& 0.33( 41)& 0.76( 69)& 0.74( 66)& 0.97( 74)& 0.61( 12)\\
& 100 &  0.08(0.5)& 0.42( 15)& 0.67( 29)& 0.58(9.1)& & 0.88(6.9)& 2.15( 79)& 1.19( 53)& 0.72( 32)& 0.82( 42)& 0.80( 12)\\
& 200 &  0.03(0.1)& 0.02(0.2)& 0.67(6.6)& 0.77(9.3)& & 0.78(2.5)&-2.84( 26)&-0.58(5.8)& 0.48(4.8)& 0.93(9.6)& 1.11( 13)\\
& 500 & -0.06(0.1)&-0.27(0.7)& 1.12(3.1)& 1.18( 10)& & 0.84(1.0)& 27.7( 74)& 23.4( 64)& 0.62(1.7)& 1.50(4.2)& 1.59( 13)\\
&1000 & -0.11(0.1)& 5.78(6.9)& 1.47(1.8)& 1.45(9.8)& &-0.36(0.2)&-63.1( 75)&-63.6( 77)& 2.48(3.0)& 2.10(2.6)& 2.04( 13)\\
&2000 & -0.52(0.1)&-25.9( 14)& 1.99(1.1)& 1.69(9.0)& & 0.36(0.1)& 0.34(0.2)& 0.94(0.5)&-13.3(7.5)& 2.82(1.6)& 2.80( 14)\\ \hline
& MAE &  0.10(5.2)& 2.29( 12)& 0.60( 14)& 0.51(9.1)& & 0.42( 16)& 6.55( 25)& 6.17( 23)& 1.35( 13)& 0.80( 16)& 0.71( 11)\\ \hline \hline
\end{tabular}}
 \begin{tablenotes}
\scriptsize
  \item[a] Reference values used to deduce basis set errors are from Ref.\ \cite{Ivanov}.
For the 1s$^2$2s$^2$ state of the B$^+$ ion, and the 1s$^2$2s$^2$2p$_{-1}$ state of the B atom,
reference values are from our own calculation by using very large basis sets 
including multiple sequences \cite{ZhuZhaTri-PRA-2014}. See text for details.
 \end{tablenotes}
\end{threeparttable}
\end{table}

\begin{table}
\caption{\label{tbl7prim} Basis set errors for the C atom in ${\bf B}$ fields$^a$.  (Absolute errors in mH, numbers in parenthesis are the relative errors in millionths, and $\mathbf B$ in a.u.)}
\begin{threeparttable}
{\scriptsize
\begin{tabular}{llllllllllllll} \hline  \hline
&       & 1s$^2$2s$^2$\hspace{2em} &1s$^2$2s & 1s$^2$2s &1s$^2$2p$_0$2p$_{-1}$ &1s$^2$2p$_{-1}$3d$_{-2}$ &1s$^2$2p$_0$2p$_{-1}$ &1s$^2$2p$_{-1}$3d$_{-2}$ \\
& $B$(a.u.) &2p$_0$2p$_{-1}$ & 2p$_0$2p$_{-1}$2p$_1$ & 2p$_0$2p$_{-1}$3d$_{-2}$ & 3d$_{-2}$4f$_{-3}$ &4f$_{-3}$5f$_{-4}$ & 4f$_{-3}$5f$_{-4}$ & 4f$_{-3}$5g$_{-4}$6h$_{-5}$ \\ \hline
&   0 &  0.35( 9.4) &   0.04( 1.2) &   0.35( 9.5) &   0.18( 5.0) &   0.22( 6.1) &   0.21( 8.7) &   0.33(  14) \\
& 0.1 &  0.50(  13) &   0.18( 4.8) &   0.55(  15) &   0.38(  10) &   0.34( 9.5) &   0.20( 8.0) &   0.48(  20) \\
& 0.2 &  0.77(  20) &   0.06( 1.6) &   0.81(  21) &   0.16( 4.5) &   0.31( 8.5) &   0.18( 7.2) &   0.41(  17) \\
& 0.5 &  0.56(  15) &   0.32( 8.4) &   0.66(  17) &   0.32( 8.5) &   0.32( 8.7) &   0.37(  14) &   0.38(  15) \\
&   1 &  0.41(  11) &   0.38( 9.8) &   0.38( 9.6) &   0.32( 8.4) &   0.30( 8.0) &   0.42(  15) &   0.37(  14) \\
&   2 &  0.54(  14) &   0.40(  10) &   0.38( 9.4) &   0.26( 6.5) &   0.40(  10) &   0.49(  16) &   0.38(  13) \\
&   5 &  0.89(  25) &   0.46(  12) &   0.64(  15) &   0.39( 9.3) &   0.45(  11) &   1.36(  38) &   0.62(  18) \\
&  10 &  0.56(  18) &   0.56(  16) &   0.43( 9.9) &   0.24( 5.5) &   0.73(  16) &   0.52(  12) &   0.60(  14) \\
&  20 &  0.28(  14) &   0.18( 7.1) &   0.43( 9.9) &   0.46(  10) &   1.08(  24) &   0.60(  12) &   0.61(  12) \\
&  50 &  1.12(  52) &   0.68(  48) &   1.02(  31) &   0.95(  26) &   1.27(  34) &   1.07(  15) &   1.39(  19) \\
& 100 &  1.27(  12) &   0.96(  10) &   0.95( 199) &   0.70(  79) &   1.63( 143) &   1.22(  14) &   1.51(  16) \\
& 200 &  1.29( 4.6) &   1.27( 4.9) &   1.25(  19) &   5.71(  94) &   1.73(  30) &   3.68(  32) &   2.08(  17) \\
& 500 &  2.22( 2.7) &   2.15( 2.7) &   1.84( 5.8) &   3.25(  11) &   2.37( 7.9) &   3.53(  22) &   3.11(  18) \\
&1000 &  2.23( 1.2) &   2.12( 1.2) &   1.77( 2.3) &   0.72( 1.0) &   3.15( 4.2) &  -3.52(  17) &   3.29(  15) \\
&2000 &  1.14( 0.3) &   1.85( 0.5) &   1.54( 0.9) &   0.90( 0.5) &   4.89( 2.9) &    ---( ---)$^b$ &   4.32(  16) \\  \hline
&MAE  &  0.94(  14) &   0.78( 9.3) &   0.87(  25) &   1.00(  19) &   1.28(  22) &   1.24(  16) &   1.33(  16) \\  \hline  \hline
\end{tabular}}
 \begin{tablenotes}
\scriptsize
  \item[a] Reference values used to deduce basis set errors are from Ref.\ \cite{Ivanov}.
  \item[b] Convergence was not reached.
 \end{tablenotes}
\end{threeparttable}
\end{table}

\newpage

\end{document}